# The virial theorem and the kinetic energy of particles of a macroscopic system in the general field concept

**Sergey G. Fedosin**

Sviazeva Str. 22-79, Perm, 614088, Perm Krai, Russian Federation

e-mail intelli@list.ru

The virial theorem is considered for a system of randomly moving particles that are tightly bound to each other by the gravitational and electromagnetic fields, acceleration field and pressure field. The kinetic energy of the particles of this system is estimated by three methods, and the ratio of the kinetic energy to the absolute value of the energy of forces, binding the particles, is determined, which is approximately equal to 0.6. For simple systems in classical mechanics, this ratio equals 0.5. The difference between these ratios arises by the consideration of the pressure field and acceleration field inside the bodies, which make additional contribution to the acceleration of the particles. It is found that the total time derivative of the system's virial is not equal to zero, as is assumed in classical mechanics for systems with potential fields. This is due to the fact that although the partial time derivative of the virial for stationary systems tends to zero, but in real bodies the virial also depends on the coordinates and the convective derivative of the virial, as part of the total time derivative inside the body, is not equal to zero. It is shown that the convective derivative is also necessary for correct description of the equations of motion of particles.

### 1. Introduction

The virial theorem relates the kinetic and potential energies of a stationary system in nonrelativistic mechanics and is widely used in astrophysics for approximate evaluation of the mass of large space systems, based on their sizes and the distribution of velocities of individual objects [1]. The theorem statement in addition to the gravitational field can also include other fields, such as the electromagnetic field and the pressure field [2]. The relativistic modification of the theorem takes into account the fact that the definition of momentum and kinetic energy of each particle of the system also includes the corresponding

Lorentz factor. In [3] by means of the virial theorem the kinetic energy in a tensor form is associated at the microscopic level with the stress tensor (Eshelby stress) in order to take into account the pressure effects within the framework of classical physics and in [4] the similar approach is used in variable-mass systems, where the fluxes of mass and energy are taken into consideration.

In contrast to this we will analyze the virial theorem for a system of closely interacting particles, which are bound to each other by the gravitational and electromagnetic fields. In this case we will use the concept of the vector pressure field, as well as the concept of the acceleration vector field, in which the role of the stress-energy tensor of matter is played by the stress- energy tensor of the acceleration field [5, 6]. All these fields are parts of the general field [7], which can be decomposed into two main components. The source of the first component is the mass four-current $J^{\mu}$, which generates such vector fields as the gravitational field, acceleration field, pressure field, dissipation field and macroscopic fields of strong and weak interactions. The second component of the general field is the electromagnetic field, the source of which is the charge four-current $j^{\mu}$.

In the derivation of the virial theorem it is generally assumed that the time derivative of the virial of the system, averaged over time, tends to zero. By direct calculation we will show that in our model this is not exactly so and will provide our explanation for this state of things, taking into account the relation between the general field components.

## 2. The virial theorem

Suppose there is a bounded system of a number of randomly moving small particles, which has a spherical shape and is in equilibrium under the action of the proper gravitational and electromagnetic fields, acceleration field and pressure field. If the spaces between the particles are small, as in a liquid, it can be assumed that the matter inside this sphere is distributed uniformly. We studied such a physical system in [8], where the field strengths, potentials and energies of all the four fields were first defined in the framework of the relativistic uniform model.

Let us place the origin of the coordinate system at the center of the sphere and apply the virial theorem to this system of particles and fields. This system is stable, the particles are bound by the forces arising from the action of the fields, and therefore the conditions of the theorem are satisfied. The virial theorem in a relativistic form can be written as follows:

$$2\langle W_k \rangle = \left\langle \frac{dG_V}{dt} \right\rangle - \left\langle \sum_{i=1}^{N} \mathbf{F}_i \cdot \mathbf{r}_i \right\rangle, \qquad (1)$$

where $G_V = \sum_{i=1}^{N}(\mathbf{p}_i \cdot \mathbf{r}_i)$ is the virial as a certain scalar function; the symbol $\langle \ \rangle$ denotes averaging over a sufficiently large period of time; $\langle W_k \rangle = \frac{1}{2}\left\langle \sum_{i=1}^{N} \gamma_i m_i v_i^2 \right\rangle$ is a quantity that in the limit of low velocities tends to the total kinetic energy of all $N$ particles of the system; $\gamma_i$ is the Lorentz factor of the $i$-th particle with the mass $m_i$ and velocity $\mathbf{v}_i$; the vectors $\mathbf{r}_i$ and $\mathbf{p}_i = \gamma_i m_i \mathbf{v}_i$ denote the radius-vector and relativistic momentum of the $i$-th particle, $\mathbf{F}_i = \frac{d\mathbf{p}_i}{dt}$ is the total force, acting on the $i$-th particle.

In order to pass on to the virial theorem in its classic form, it is sufficient to equate in (1) the Lorentz factors of all the particles to unity, that is to assume $\gamma_i = 1$.

Figure 1 shows that in contrast to the discrete distribution of particles, in order to use the continuous distribution approximation it is necessary to detect in the matter of the system under consideration the particles of such a size, that the gaps between them would be close to zero, and to assume that the sphere's volume is made up of the volumes of these particles. Each particle of this type occupies a certain representative volume element of the system, which is sufficient for the correct description of the typical properties of particles and acting fields.

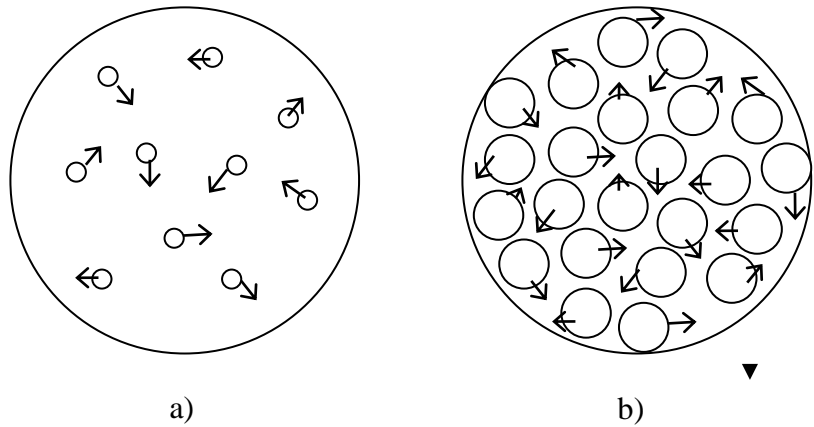

a)               b)

**Figure 1.** a) The system of a spherical form, containing randomly moving discrete particles of a small size. b) Approximation of the continuous distribution of matter with closely interacting particles. The arrows show the directions of instantaneous velocities of the particles.

Let us first calculate the value $2\langle W_k \rangle$ in the left-hand side of (1), substituting $m_i$ with $dm = \rho_0 \gamma' dV$ and the sum over all the particles with the integral over the volume of the fixed sphere. The value $\rho_0$ is the mass density in the reference frames associated with the particles, $\gamma'$ is the Lorentz factor of the moving particles, the product $\rho_0 \gamma'$ gives the mass density of the particles from the point of view of the observer, stationary with respect to the sphere, and the volume element $dV$ inside the sphere corresponds to the volume of a particle from the point of view of this observer. We can assume that the total volume of the particles at rest is greater than the volume of the sphere, but because of the motion the volume of each particle decreases due to the length contraction effect of the special theory of relativity.

Actually, to apply the relativistic formulas correctly, the volume of any moving spherical particle is modeled by the so-called Heaviside ellipsoid, which has been first mentioned in [9]. It turns out that the volume of this ellipsoid is smaller than the volume of the particle under consideration in the reference frame associated with the particle by the value of the Lorentz factor. All this leads to the fact that the total volume of the particles moving inside the sphere becomes equal to the volume of the sphere.

The same can be said in other words. If we divide the system's matter into separate independently moving particles, as is shown in Fig. 1 b, then the sum of the volumes of Heaviside ellipsoids of all particles should be equal to the volume of the sphere and the sum of the proper volumes of these particles, respectively, will be greater than the volume of the sphere.

According to [8], the Lorentz factor $\gamma'$ for the particles inside the fixed sphere is a function of the current radius $r$:

$$\gamma' = \frac{c \gamma_c}{r\sqrt{4\pi \eta \rho_0}} \sin\left(\frac{r}{c}\sqrt{4\pi \eta \rho_0}\right) \approx \gamma_c - \frac{2\pi \eta \rho_0 r^2 \gamma_c}{3c^2}, \tag{2}$$

where $c$ is the speed of light, $\eta$ is the acceleration field coefficient, $\gamma_c = \dfrac{1}{\sqrt{1-v_c^2/c^2}}$ is the Lorentz factor for the velocities $v_c$ of the particles at the center of the sphere, and in view of the smallness of the argument the sine can be expanded into second-order terms.

The second expansion term in (2) can be represented as follows:

$$\frac{2\pi \eta \rho_0 r^2 \gamma_c}{3c^2} = \frac{\eta G m(r) \gamma_c}{2Grc^2} = -\frac{\eta \psi(r) \gamma_c}{2Gc^2},$$

where the expression $m(r) = \dfrac{4\pi \rho_0 r^3}{3}$ gives an estimate of the mass contained inside the current radius $r$ of the sphere, $\psi(r) = -\dfrac{Gm(r)}{r}$ is the gravitational potential created by the spherical mass $m(r)$ on the radius $r$.

In cosmic bodies held by their own gravitation the acceleration field coefficient $\eta$ differs from the gravitational constant $G$ only by a small numerical factor of the order of unity, and the Lorentz factor $\gamma_c$ is only slightly greater than unity for the majority of bodies. As a result, the second expansion term in (2) can be considered as the ratio of the absolute value of the average gravitational potential inside the body to the square of the speed of light. This ratio is small and starts to increase significantly only in white dwarfs and neutron stars. Despite the smallness of the second term, it is absolutely essential for justification of our relativistic approach. Indeed, let us square the equation for $\gamma'$ in (2) and we will obtain approximately the following:

$$v'^2 \approx v_c^2 - \frac{4\pi \eta \rho_0 r^2}{3}. \tag{3}$$

The velocity $v'$ of the random motion of particles inside the sphere is a function of the current radius $r$ only. In this case, as the volume element we can take the volume of the thin spherical layer: $dV = 4\pi r^2 dr$. Equating $\gamma_i$ and $v_i^2$ from (1) to $\gamma'$ in (2) and $v'^2$ in (3), respectively, and substituting $m_i$ with $dm = \rho_0 \gamma' dV$, for $2\langle W_k \rangle$ we obtain the following:

$$2\langle W_k \rangle = \left\langle \sum_{i=1}^{N} \gamma_i m_i v_i^2 \right\rangle \approx 4\pi \rho_0 \int_0^a \left( \gamma_c - \frac{2\pi \eta \rho_0 r^2 \gamma_c}{3c^2} \right)^2 \left( v_c^2 - \frac{4\pi \eta \rho_0 r^2}{3} \right) r^2 dr \approx$$

$$\approx m v_c^2 \gamma_c^2 - \frac{3\eta m^2 \gamma_c^2}{5a} - \frac{3\eta m^2 v_c^2 \gamma_c^2}{5ac^2} + \frac{3\eta^2 m^3 \gamma_c^2}{7a^2 c^2}. \qquad (4)$$

Here, the mass $m$, which has auxiliary character, is equal to the product of the density $\rho_0$ by the volume of the sphere, with the radius of the sphere equal to $a$.

By analogy with [5], the equation of motion of the particles can be written as follows:

$$\rho_0 \frac{d(\gamma' \mathbf{v}')}{dt} = \rho_{0q}(\mathbf{E} + [\mathbf{v} \times \mathbf{B}]) + \rho_0 (\mathbf{\Gamma} + [\mathbf{v} \times \mathbf{\Omega}]) + \rho_0 (\mathbf{C} + [\mathbf{v} \times \mathbf{I}]). \qquad (5)$$

A distinctive feature of this equation as opposed to the equation of motion in [5] is that it is written not for a particular physical particle but for a representative particle, which behaves as a certain typical particle averaged with respect to all parameters. This is indicated by the fact that $\gamma'$ as the Lorentz factor and the velocity $\mathbf{v}'$ are used, which are found from the equations of the acceleration field, while for a physical particle usually its instantaneous velocity and the Lorentz factor, corresponding to this velocity, are taken into account. We can assume that Eq. (5) is the result of averaging of the equations of motion for a certain ensemble of physical particles, so that a typical equation of motion for typical particles is obtained.

The right-hand side of (5) represents the total density of the force acting on a typical particle inside the sphere. This force density must be multiplied by the radius $\mathbf{r}$, where this particle is located, and then integrated over the entire volume of the sphere in order to calculate the second term in the right-hand side of (1). The force density in (5) should also be multiplied by $\gamma'$ in order to obtain the product $\rho_0 \gamma' \frac{d(\gamma \mathbf{v})}{dt}$, which in view of the expression $dm = \rho_0 \gamma' dV$ after integration over the volume will be equivalent to the expression for the force $\mathbf{F}_i = \frac{d(\gamma_i m_i \mathbf{v}_i)}{dt} = m_i \frac{d(\gamma_i \mathbf{v}_i)}{dt}$.

We will take into account that the magnetic induction vector $\mathbf{B}$, the solenoidal vector (the torsion field) $\mathbf{\Omega}$ of the gravitational field and the solenoidal vector $\mathbf{I}$ of the pressure field inside the fixed sphere are equal to zero due to the random motion of particles. In this case, according to [8] we have the following:

$$\mathbf{E} = \frac{\gamma_c \rho_{0q} \mathbf{r}}{3\varepsilon_0}, \qquad \mathbf{\Gamma} = -\frac{4\pi G \gamma_c \rho_0 \mathbf{r}}{3}, \qquad \mathbf{C} = \frac{4\pi \sigma \gamma_c \rho_0 \mathbf{r}}{3},$$

and we can write:

$$-\left\langle \sum_{i=1}^{N} \mathbf{F}_i \cdot \mathbf{r}_i \right\rangle = -\int_V \gamma'(\rho_{0q} \mathbf{E} + \rho_0 \mathbf{\Gamma} + \rho_0 \mathbf{C}) \cdot \mathbf{r}\, dV =$$
$$= -\frac{16\pi^2 \rho_0^2 \gamma_c^2}{3} \int_0^a \left(1 - \frac{2\pi \eta \rho_0 r^2}{3c^2}\right)\left(\frac{\rho_{0q}^2}{4\pi\varepsilon_0 \rho_0^2} - G + \sigma\right) r^4\, dr = \qquad (6)$$
$$= \frac{3\eta m^2 \gamma_c^2}{5a} - \frac{3\eta^2 m^3 \gamma_c^2}{14 a^2 c^2}.$$

In (5) and (6), $\mathbf{E}$, $\mathbf{\Gamma}$ and $\mathbf{C}$ are the field strengths of the electric field, the gravitational field and the pressure field, respectively, $\rho_{0q}$ is the charge density of the particles in the reference frames associated with the particles, $\varepsilon_0$ is the vacuum permittivity, $G$ is the gravitational constant, and $\sigma$ is the pressure field coefficient. Besides, the correlation between the field coefficients was used, which had been obtained in [10] using the equations of motion:

$$\eta + \sigma = G - \frac{\rho_{0q}^2}{4\pi\varepsilon_0 \rho_0^2}. \qquad (7)$$

In order to arrive at (7), it will suffice to express the left-hand side of (5) in terms of the field strength $\mathbf{S}$ and the solenoidal vector $\mathbf{N}$ of the acceleration field in the framework of the special theory of relativity [5]:

$$\rho_0 \frac{d(\gamma' \mathbf{v}')}{dt} = -\rho_0 \bigl(\mathbf{S} + [\mathbf{v}' \times \mathbf{N}]\bigr).$$

Next we should take into account the equality of the solenoidal vector $\mathbf{N}$ and the vector potential $\mathbf{U}$ to zero in the system under consideration, as well as the expression for the scalar potential of the acceleration field in the form $\vartheta = c^2 \gamma'$. Then the field strength of the acceleration field in view of (2) is given by the formula:

$$\mathbf{S} = -\nabla \vartheta - \frac{\partial \mathbf{U}}{\partial t} = \frac{4\pi \eta \gamma_c \rho_0 \mathbf{r}}{3}.$$

Using further in (5) the expressions given above for $\mathbf{E}$, $\mathbf{\Gamma}$ and $\mathbf{C}$, we arrive at (7). Actually relation (7) for the fields' coefficients is the consequence of the local balance of the forces and energies, associated with the fields, acting on the particles.

The virial $G_V = \sum_{i=1}^{N} (\gamma_i m_i \mathbf{v}_i \cdot \mathbf{r}_i)$ contains the scalar vector products of the form $\mathbf{v}_i \cdot \mathbf{r}_i$. In these products we will substitute the particles' velocities $\mathbf{v}_i$ with the averaged velocities of random motion $\mathbf{v}'$, which depend on the current radius, according to (3). Then we will assume that $\mathbf{v}' = \mathbf{v}_r + \mathbf{v}_\perp$, where $\mathbf{v}_r$ denotes the averaged velocity component, directed along the radius, and $\mathbf{v}_\perp$ is the averaged velocity component, perpendicular to the current radius. Then for the particles inside the sphere $\mathbf{v}_i \cdot \mathbf{r}_i = \mathbf{v}_r \cdot \mathbf{r}$. Based on statistical considerations, it follows that:

$$v'^2 = v_r^2 + v_\perp^2 = 3v_r^2. \tag{8}$$

For the dependence of the magnitude of the radial velocity component on the current radius we can write in the first approximation:

$$v_r \approx A(1 - Br^2). \tag{9}$$

Thus, we assume that the dependence of the radial velocity component $v_r$ on the radius due to its form can be represented similarly to the squared velocity $v'^2$ in (3). To prove this assumption we will square expression (9) and substitute it in (8) instead of $v_r^2$, and then we will find the value of $v'^2$ and compare it with (3).

This allows us to estimate the coefficients $A$ and $B$ and to rewrite (9) as follows:

$$v_r \approx \frac{v_c}{\sqrt{3}}\left(1 - \frac{2\pi\eta\rho_0 r^2}{3v_c^2}\right). \tag{10}$$

Therefore, for the product of vectors in the virial we will have approximately the following:

$$\mathbf{v}_i \cdot \mathbf{r}_i = \mathbf{v}_r \cdot \mathbf{r} \approx \frac{v_c r}{\sqrt{3}} - \frac{2\pi\sqrt{3}\eta\rho_0 r^3}{9v_c}. \tag{11}$$

The time derivative of the virial in (1) should be regarded as the material derivative:

$$\frac{dG_V}{dt} = \frac{\partial G_V}{\partial t} + \mathbf{v}\cdot\nabla G_V, \tag{12}$$

besides, in our case the virial does not depend on time and $\frac{\partial G_V}{\partial t} = 0$.

We can calculate the product $\mathbf{v}\cdot\nabla G_V$, substituting in it $\mathbf{v}$ with $\mathbf{v}_r$, since the virial $G_V$ depends only on the radius, and the virial gradient $\nabla G_V$ is directed along the radius. Taking into account (11), (2) for the Lorentz factor, (10) for the magnitude of the radial velocity $v_r$, as well the expression $dm = \rho_0 \gamma' dV = 4\pi\rho_0 \gamma' r^2 dr$ used instead of the mass $m_i$, which before that must be taken outside the gradient sign, we find:

$$\mathbf{v}\cdot\nabla G_V = \left\langle \sum_{i=1}^{N} \mathbf{v}\cdot\nabla(\gamma_i m_i \mathbf{v}_i \cdot \mathbf{r}_i) \right\rangle \approx$$

$$\approx 4\pi\rho_0 \int_0^a \left(\gamma_c - \frac{2\pi\eta\rho_0 r^2 \gamma_c}{3c^2}\right) \mathbf{v}_r \cdot \nabla\left[\left(\gamma_c - \frac{2\pi\eta\rho_0 r^2 \gamma_c}{3c^2}\right)\left(\frac{v_c r}{\sqrt{3}} - \frac{2\pi\sqrt{3}\eta\rho_0 r^3}{9v_c}\right)\right] r^2 dr \approx$$

$$\approx \frac{mv_c^2 \gamma_c^2}{3} - \frac{2\eta m^2 v_c^2 \gamma_c^2}{5ac^2} - \frac{2\eta m^2 \gamma_c^2}{5a} + \frac{3\eta^2 m^3 \gamma_c^2}{7a^2 c^2} + \frac{3\eta^2 m^3 \gamma_c^2}{28a^2 v_c^2} - \frac{\eta^3 m^4 \gamma_c^2}{9a^3 v_c^2 c^2}.$$

$$\tag{13}$$

Substituting (13) into (12), provided that $\frac{\partial G_V}{\partial t} = 0$, finding $\frac{dG_V}{dt}$ and using it in (1), in view of (4) and (6) we obtain an approximate relation:

$$v_c^2 \approx \frac{6\eta m}{5a} + \frac{9\eta^2 m^2}{56 a^2 v_c^2}.$$

Considering this relation as a quadratic equation for $v_c^2$ and solving this equation, we arrive at the following:

$$v_c^2 \approx \frac{3\eta m}{5a}\left(1+\frac{9}{\sqrt{56}}\right) \approx \frac{1.3216\eta m}{a}. \qquad (14)$$

With $r=a$ and in view of (14), Eq. (3) gives an expression for the squared velocity of the particles near the sphere's surface:

$$v_s^2 = v_c^2 - \frac{\eta m}{a} \approx \frac{0.3216\eta m}{a}.$$

Hence it follows that $\frac{v_c^2}{v_s^2} \approx 4.1$. This means that as a consequence of the virial theorem, the square of the velocity of particles $v_c^2$ in the center is about 4 times greater than the square of the velocity of particles $v_s^2$ near the surface of the sphere. Since the squared velocities are proportional to the kinetic energy and temperature, then, under condition of the constant mass density $\rho_0$ in the considered idealized system, the temperatures at the center and near the surface must not differ more than 4.1 times. Among the real objects, the density of which does not change much with the current radius, we can take Bok globules. Their typical radius is 0.35 parsecs, the mass is 11 Solar masses, and the recorded temperature of dust in some globules may reach 26 K [11]. In [10], based on the equations of motion of particles, the kinetic temperature of the particles near the surface of a globule was estimated: $T_s = 5.5$ K. If we assume that $T_c = 4.1 T_s$, then the central temperature is $T_c = 22.5$ K, which is close enough to observations.

Using (14) for substituting in (4) and in (13) in view of (12):

$$2\langle W_k \rangle \approx \frac{0.7216\eta m^2 \gamma_c^2}{a}, \qquad \frac{dG_V}{dt} \approx \mathbf{v}\cdot\nabla G_V \approx \frac{0.1216\eta m^2 \gamma_c^2}{a}. \qquad (15)$$

Expressions (15) and (6) agree well with the virial theorem (1) in the approximation under consideration. In addition, in (15) we can see that the time derivative of the virial $\frac{dG_V}{dt}$ after averaging is not equal to zero, as is usually assumed in classical mechanics. From (15) and (6) the relation follows:

$$\langle W_k \rangle \approx -0.6 \left\langle \sum_{i=1}^{N} \mathbf{F}_i \cdot \mathbf{r}_i \right\rangle. \qquad (16)$$

Meanwhile, in the conventional interpretation of the virial theorem the kinetic energy of the system of particles must be two times less than the energy, associated with the forces, holding the particles:

$$\langle W_k \rangle_m = -0.5 \left\langle \sum_{i=1}^{N} \mathbf{F}_i \cdot \mathbf{r}_i \right\rangle, \qquad (17)$$

If we substitute (14) in (10), we will obtain the following:

$$v_r \approx \frac{v_c}{\sqrt{3}}\left(1 - \frac{2\pi\eta\rho_0 r^2}{3v_c^2}\right) \approx 0.6637\sqrt{\frac{\eta m}{a}}\left(1 - \frac{0.3783 r^2}{a^2}\right).$$

If we take into account (8), then for the velocities' amplitudes we can write $v_\perp = \sqrt{2}v_r$. We see that inside the sphere there are radial gradients both of the radial component $\mathbf{v}_r$ and of the velocity component $\mathbf{v}_\perp$ perpendicular to the radius, and also there is a gradient of the squared velocity of particles $v'^2$ in (3), while $\mathbf{v}' = \mathbf{v}_r + \mathbf{v}_\perp$. The velocity $\mathbf{v}_\perp$ leads to a certain centripetal acceleration directed along the radius. Due to this as well as due to the radial action of the pressure field and the electric field the acceleration arises, which counteracts the gravitational acceleration and leads to a noticeable difference between (16) and (17).

The difference of (16) from the classical case (17) is caused by the fact that we take into account not the usual uniformity of mass and charge in the reference frame of the sphere, but the relativistic uniformity, when the mass and charge densities are constant in their own reference frames, associated with individual particles. This leads to a change in the values of the field strengths of all the fields inside the sphere and of the field strengths of the gravitational and electromagnetic fields outside the sphere, as well as to a change in accelerations from the action of respective forces.

It follows from (16) that at the constant potential energy (6), associated with the forces holding the particles of the system, the kinetic energy of motion must be greater than in (17) by a value of about 20%. As the density non-uniformity inside the system increases, the difference between (16) and (17) may change even more.

### 3. The kinetic energy: standard definition

Within the framework of the special theory of relativity, the kinetic energy of a particle is calculated as the difference between the relativistic energy of a moving particle and the energy of the particle at rest. For a system of $N$ particles we obtain the following:

$$E_k = \sum_{i=1}^{N} (\gamma_i - 1) m_i c^2 . \qquad (18)$$

Let us use in (18) instead of $\gamma_i$ the Lorentz factor $\gamma'$ from (2), and replace the mass $m_i$ by $dm = \rho_0 \gamma' dV = 4\pi \rho_0 \gamma' r^2 dr$ and the sum over all the particles by the integral over the volume of a fixed sphere:

$$E_k = 4\pi \rho_0 c^2 \int_0^a [\gamma' - 1] \gamma' r^2 dr .$$

$$E_k = \frac{4\pi \rho_0 c^3 \gamma_c}{\sqrt{4\pi \eta \rho_0}} \int_0^a \left[ \frac{c \gamma_c}{r \sqrt{4\pi \eta \rho_0}} \sin\left(\frac{r}{c}\sqrt{4\pi \eta \rho_0}\right) - 1 \right] \sin\left(\frac{r}{c}\sqrt{4\pi \eta \rho_0}\right) r \, dr \approx$$
$$\approx mc^2 \gamma_c^2 - \frac{3\eta m^2 \gamma_c^2}{5a} - mc^2 \gamma_c + \frac{3\eta m^2 \gamma_c}{10a} . \qquad (19)$$

Let us substitute the Lorentz factor in (19):

$$\gamma_c = \frac{1}{\sqrt{1-v_c^2/c^2}} \approx 1 + \frac{v_c^2}{2c^2} + \frac{3v_c^4}{8c^4}, \qquad E_k \approx \frac{mv_c^2 \gamma_c}{2} - \frac{3\eta m^2 \gamma_c}{10a} - \frac{3\eta m^2 v_c^2 \gamma_c}{10ac^2} + \frac{3mv_c^4 \gamma_c}{8c^2}.$$

(20)

Substituting $v_c^2$ from (14) in (20) we find the approximate expression for the kinetic energy:

$$E_k \approx \frac{0.3608\eta m^2 \gamma_c}{a} + \frac{0.2585\eta^2 m^3 \gamma_c}{a^2 c^2}.$$

(21)

Within the limit of low velocities, with an accuracy up to the terms of the second order of smallness, the expression for $E_k$ in (21) coincides with the energy $\langle W_k \rangle$ in (15), which proves our calculations of the kinetic energy based on the relativistic energy definition and the energy estimate based on the virial theorem.

Note that when we determine $E_k$ we use the Lorentz factor $\gamma'$ from (2), which is found through the acceleration field of the particles inside the sphere, instead of the Lorentz factor of individual particles moving randomly. Thus, the kinetic energy $E_k$ in (19) and (21) is obtained as a certain approximation to the actual kinetic energy of the particles.

### 4. The energy of motion

In [5] we gave the definition of the energy of the particles' motion using the generalized 3-momenta $\mathbf{P}_n$ of the system's particles:

$$E_V = \frac{1}{2}\sum_{n=1}^{N} (\mathbf{P}_n \cdot \mathbf{v}_n),$$

where $\mathbf{v}_n$ is the 3-vector of velocity of the particle with the number $n$; $N$ specifies the number of particles in the system; and the generalized momentum of each particle is expressed in terms of the particle Lagrangian $L_n$ according to the formula: $\mathbf{P}_n = \frac{\partial L_n}{\partial \mathbf{v}_n}$.

This energy can also be written as the half-sum of the Hamiltonian $H$ and the Lagrangian $L$ of the system of particles and four fields:

$$E_V = \frac{1}{2}(H+L) = \frac{1}{2c}\sum_{n=1}^{N}\int^{n}\left(\rho_0 \mathbf{v}_n \cdot \mathbf{U}_n + \rho_0 \mathbf{v}_n \cdot \mathbf{D}_n + \rho_{0q}\mathbf{v}_n \cdot \mathbf{A}_n + \rho_0 \mathbf{v}_n \cdot \mathbf{\Pi}_n\right) u^0 \sqrt{-g}\, dx^1 dx^2 dx^3.$$

(22)

where $\mathbf{U}_n$, $\mathbf{D}_n$, $\mathbf{A}_n$ and $\mathbf{\Pi}_n$ denote the vector potentials of the acceleration field, gravitational field, electromagnetic field and pressure field, respectively; $u^0$ is the time component of the particle's four-velocity; $g$ is the determinant of the metric tensor; $dx^1 dx^2 dx^3$ is the product of the spatial coordinates' differentials.

The Hamiltonian $H$ and the Lagrangian $L$ of the system, which are present in (22), were determined in [5] in a covariant way for the curved spacetime, while the expression for the conserved over time relativistic energy of an arbitrary isolated system coincides with the expression for the Hamiltonian $H$. It should be noted that the energy of motion $E_V$ does not contain any scalar curvature or the cosmological constant and thus does not depend on the method of gauging the relativistic energy of the system.

Within the framework of the special theory of relativity for the particles inside a fixed sphere we can assume that $u^0 = c\gamma'$. In the random motion of particles the total vector potentials of all the fields, averaged over the entire set of particles, are equal to zero. However, the vector potentials of each individual particle are equal to zero only at rest, but in case of motion they are proportional to the particle velocity and to the scalar potentials of the proper fields of particles and inversely proportional to the squared speed of light. This follows from the definition of the four-potential of each field [6], as well as from the solution of the wave equation for the vector potential of the corresponding field at a constant velocity of the particle's motion.

It should be noted that in (22) integration is done over the volume of each particle separately and then summation is performed over all the particles. In integration over the volume of one particle, the velocity of this particle is considered as a constant and can be taken out of the integral sign. As a result, we can rewrite (22) as follows:

$$E_V = \sum_{n=1}^{N}\frac{v_n^2}{2c^2}\int^{n}\left(\rho_0 \vartheta_n + \rho_0 \psi_n + \rho_{0q}\varphi_n + \rho_0 \wp_n\right)\gamma'\, dx^1 dx^2 dx^3. \qquad (23)$$

In (23) $\vartheta_n$, $\psi_n$, $\varphi_n$ and $\wp_n$ denote the proper scalar potentials of the moving particles for the acceleration field, gravitational field, electromagnetic field and pressure field, respectively; the quantity $\gamma'$ is the Lorentz factor of particles according to (2). As it was shown in [12], from the gauge of the system energy using the cosmological constant $\Lambda$ the following expression was obtained:

$$\sum_{n=1}^{N}\int_{}^{n}\left(\rho_0\vartheta_n+\rho_0\psi_n+\rho_{0q}\varphi_n+\rho_0\wp_n\right)\gamma'\,dx^1\,dx^2\,dx^3 = \sum_{n=1}^{N}M_{0n}c^2 = M_0 c^2 = -\int ck\Lambda\,dx^1\,dx^2\,dx^3,$$

where $k = -\dfrac{c^3}{16\pi G\beta}$; $\beta$ is the constant of the order of unity, which is included as a multiplier in the equation for the metric; $M_0 = \sum_{n=1}^{N} M_{0n}$ is the gauge mass as the total mass of the system's particles, removed from the system to infinity and being there at rest, taking into account the energies of particles in the potentials of the proper fields, but neglecting the field energies as such.

From (23) we then obtain the following:

$$E_V = \frac{1}{2}\sum_{n=1}^{N} M_{0n} v_n^2. \tag{24}$$

Comparison with the expression $\langle W_k \rangle = \dfrac{1}{2}\left\langle \sum_{i=1}^{N} \gamma_i m_i v_i^2 \right\rangle$ from (1) shows that $\langle W_k \rangle$ tends to the energy $E_V$ only in the limit of low velocities, where we can neglect the Lorentz factors of the particles.

As a first approximation, in (24) we will replace the mass $M_{0n}$ by $dm = \rho_0 \gamma' dV = 4\pi \rho_0 \gamma' r^2 dr$, the squared velocity $v_n^2$ by $v'^2$ from (3), use $\gamma'$ from (2), and represent the sum over all the particles as the integral over the volume of the fixed sphere:

$$E_V \approx 2\pi \rho_0 \int_0^a v'^2 \gamma' r^2\,dr = \frac{mv_c^2 \gamma_c}{2} - \frac{3\eta m^2 \gamma_c}{10a} - \frac{3\eta m^2 v_c^2 \gamma_c}{20ac^2} + \frac{3\eta^2 m^3 \gamma_c}{28a^2 c^2}.$$

If we equate the obtained integral for $E_V$ to the first integral for $E_k$ from (19), it gives the relation for the Lorentz factor of the form $\gamma' \approx 1 + \frac{v'^2}{2c^2}$, which can be considered valid in the first-order approximation. The difference between $E_V$ and $E_k$ in (20) occurs in the terms containing the squared speed of light in the denominator. Hence it follows that the energy of motion (22), determined with the help of the generalized momenta and the proper fields of particles, is close enough to the kinetic energy of particles $E_k$, which is found with the help of the distribution of particles in the acceleration field.

### 5. The analysis of the equation of motion

We will transform the equation of motion of typical particles (5) multiplying it scalarly by the vector quantity $\gamma' \mathbf{v}'$:

$$\rho_0 \gamma' \mathbf{v}' \frac{d(\gamma' \mathbf{v}')}{dt} = \frac{\rho_0}{2} \frac{d(\gamma'^2 v'^2)}{dt} = \rho_{0q} \gamma' \mathbf{E} \cdot \mathbf{v}' + \rho_0 \gamma' \mathbf{\Gamma} \cdot \mathbf{v}' + \rho_0 \gamma' \mathbf{C} \cdot \mathbf{v}'. \qquad (25)$$

We will calculate the term on the left-hand side of this equation, substituting the Lorentz factor $\gamma'$ from (2), and the value of $v'^2$ from (3). In this case the total time derivative will be regarded as the material derivative, in which in the convective derivative the velocity $\mathbf{v}_r$ with amplitude (10) can be used instead of the velocity $\mathbf{v}$. Neglecting the small terms with the square of the speed of light, we find:

$$\frac{\rho_0}{2} \frac{d(\gamma'^2 v'^2)}{dt} = \frac{\rho_0}{2} \frac{\partial(\gamma'^2 v'^2)}{\partial t} + \frac{\rho_0}{2} \mathbf{v}_r \cdot \nabla(\gamma'^2 v'^2) \approx$$

$$\approx \frac{\rho_0 v_c}{2\sqrt{3}} \left(1 - \frac{2\pi \eta \rho_0 r^2}{3 v_c^2}\right) \frac{\partial}{\partial r}\left[\left(\gamma_c - \frac{2\pi \eta \rho_0 r^2 \gamma_c}{3c^2}\right)^2 \left(v_c^2 - \frac{4\pi \eta \rho_0 r^2}{3}\right)\right] \approx \qquad (26)$$

$$\approx -\frac{4\pi \eta \rho_0^2 v_c \gamma_c^2 r}{3\sqrt{3}} \left(1 - \frac{2\pi \eta \rho_0 r^2}{3 v_c^2}\right).$$

Since the system under consideration is stationary, then the partial time derivative $\frac{\partial(\gamma'^2 v'^2)}{\partial t}$ in (26) is equal to zero. The need to use the material derivative in (25) and (26) is due to the fact that the product $\gamma'^2 v'^2$ is the function of the spatial coordinates, but is time-independent. However, relation (25) must be valid for all reference frames, including the reference frame moving radially at the velocity $\mathbf{v}_r$. In this reference frame the gradient $\nabla(\gamma'^2 v'^2)$ is other than zero, and then the derivative $\dfrac{d(\gamma'^2 v'^2)}{dt}$ is not equal zero.

We will now calculate the right-hand side of (25), substituting there the expressions for the field strengths $\mathbf{E}$, $\mathbf{\Gamma}$ and $\mathbf{C}$, which were used in (6). In this case we will take into account that all the forces are directed along the radius and therefore the velocity $\mathbf{v}'$ can be replaced with $\mathbf{v}_r$:

$$\rho_{0q} \gamma' \mathbf{E} \cdot \mathbf{v}' + \rho_0 \gamma' \mathbf{\Gamma} \cdot \mathbf{v}' + \rho_0 \gamma' \mathbf{C} \cdot \mathbf{v}' =$$
$$= \frac{v_c}{\sqrt{3}} \left(1 - \frac{2\pi \eta \rho_0 r^2}{3 v_c^2}\right) \left(\gamma_c - \frac{2\pi \eta \rho_0 r^2 \gamma_c}{3 c^2}\right) \left(\frac{\gamma_c \rho_{0q}^2 r}{3\varepsilon_0} - \frac{4\pi G \gamma_c \rho_0^2 r}{3} + \frac{4\pi \sigma \gamma_c \rho_0^2 r}{3}\right) \approx \quad (27)$$
$$\approx -\frac{4\pi \eta \rho_0^2 v_c \gamma_c^2 r}{3\sqrt{3}} \left(1 - \frac{2\pi \eta \rho_0 r^2}{3 v_c^2}\right).$$

In (27), relation (7) for the field coefficients was used. Within the accuracy of the assumptions made for the velocities and the Lorentz factor, expressions (26) and (27) coincide, illustrating the validity of equation of motion (5) for the averaged velocities of particles inside the sphere and the need to use the material derivative.

### 6. Conclusion

In (1) we presented the relativistic expression of the virial theorem and then calculated each term of this expression. In (10) we obtained the approximate dependence of the amplitude of the radial component of the particles' velocity on the current radius, which is associated with the acceleration field acting in the system. For a stationary system the partial time derivative of the virial vanishes, and it becomes important to take into account the dependence of the virial on the space coordinates in the expression for the material derivative (12). As a consequence of the virial theorem, it becomes possible to estimate the velocity of particles at the center of the system in relation (14), and then to express the kinetic energy in terms of the acceleration field coefficient in (15).

In (16) we obtained the coefficient, relating the kinetic energy of particles and the energy of the forces acting on them, which is approximately equal to 0.6. Taking into account the pressure field and the acceleration field leads to 20 % difference between this coefficient and the standard value of 0.5 in (17) for systems without pressure.

From the physical standpoint, the discrepancy between these coefficients arises as a result of different interpretations of the concept of a homogeneous system: in classical mechanics the body mass density at each point is assumed to be the same in the reference frame, associated with the body, but in relativistic mechanics the mass density must be the same for each particle of the body, regardless of its motion, i.e., to be invariant under Lorentz transformations. The invariant mass density of the body's particle is the density, which is found in the reference frame associated with this particle. As a result, the particles, that move at the center of the body and have an increased velocity, have a greater mass density in the reference frame, associated with the body, which leads to the radial gradient of density and other variables inside the body in question and to correction of the virial theorem.

To check our calculations, the kinetic energy was calculated in (21) in another way, as a difference between the energies of the moving particles and the particles at rest. In (24) we also estimated the energy of the particles' motion using the generalized momenta and the proper fields of the particles, which turned out to be almost exactly equal to the kinetic energy.

In (26) and (27) we also checked whether equation of motion (5) is precisely satisfied, when the expression for the radial velocity (10) is used for the case, in which not the velocity of a specific particle is substituted in the equation of motion but the averaged random velocity of particles as a function of the radius. It turns out that in this case the time derivative in the equation of motion should be regarded as material derivative, which takes into account not only the change in the velocity over time, but also the dependence of the velocity on the coordinates.

Indeed, in the stationary case the time derivatives of physical quantities are equal to zero and the angular and radial dependences of these quantities become important. In this case, in real bodies the gravitational field is counteracted by the acceleration field, pressure field and electromagnetic field. If we split the motion of particles into oscillatory motions along the radius and to motions perpendicular to the radius, then from the standpoint of the kinetic theory the radial motions lead to normal pressure, and the motions perpendicular to the radius must be accompanied by a centripetal force, which can be associated with the force from the acceleration field. Hence it follows that simple equating of the gravitational force and the pressure force in calculation of the state of matter of cosmic bodies is not well-founded, since it does not take into account the effect of the acceleration field.

**Competing Interests**

The author declares no competing interests.